# Which research in design creativity and innovation? Let us not forget the reality of companies

**Bernard Yannou**
Professor of Design & Industrial Engineering
Laboratoire Genie Industriel
Ecole Centrale Paris
Grande Voie des Vignes - 92290 Chatenay-Malabry, France
bernard.yannou@ecp.fr

**Abstract**:

Studying design creativity and innovation from practical perspectives for companies requires both a good understanding of the company ecosystem and its inner processes contributing to delivered innovations and a rigorous design research methodology to provide effective design models, methods, platforms that are truly effective in the context of company. Working in an Industrial Engineering laboratory, we advocate a more systemic vision of design creativity and innovation in company ecosystems. We present in this paper an attempt to develop and make professional an innovation engineering. Our research works are illustrated along the different research topics of an innovation process. We start by a recent survey on innovation practice and organizational models led in 28 large companies. The lessons learned about this survey reinforce our belief that there is a need for a new method in agile management of radical innovation projects in company contexts. We currently develop, test and apply such a methodology named: *Radical Innovation Design*® (RID). Its effectiveness has been evaluated through a large scale evaluation of the project outcomes for the company. Two extensions of RID have been proposed and deployed in company contexts: a selection procedure for innovation clusters and a value-driven process for airplane development projects.

## 1. Our framework: making value for companies

One may and one must study design creativity and innovation from theoretical perspectives and carefully study cognitive aspects of it within the ideation process which is the heart of the innovation process. But one must not forget realities of companies because they are the first beneficiaries of the practical perspectives so as to get effective methods in companies and to develop ever more innovative products and services.

But studying design creativity and innovation from practical perspectives for companies is not much developed today since many barriers make it uneasy. Indeed, in a company, there are many more influencing factors than in a controlled lab experiment, experiments are not repeatable, company executives do not tolerate non profitable experiments, designers are most of the time working on different projects at a time, and it is hard and time consuming to properly measure the effects of a creativity method or innovation theory.

On the other hand, another research community, let us say in strategic product innovation management (see (Astebro 2004, Millier 1999)), studies using survey data (with both distal and proximal factors) the most influencing factors to innovation success characterizing both the company ecosystem and the innovation processes itself. But, as they are more economists working outside innovation processes at a strategic level, their a posteriori surveys lost correlations between in-process factors. Consequently, they are only capable of explaining company discrepancies for contributing or not to the success of innovations. Such researches lead to continuous improvements of design organization and resources of the companies. In a design research approach, we should be able to address the issue that, given a company with its own ecosystem (its design organization, competitors, product portfolio, present clients, designers…), how must we tune the inner (endogenous) factors of a given innovative design process to increase the probability of its success.

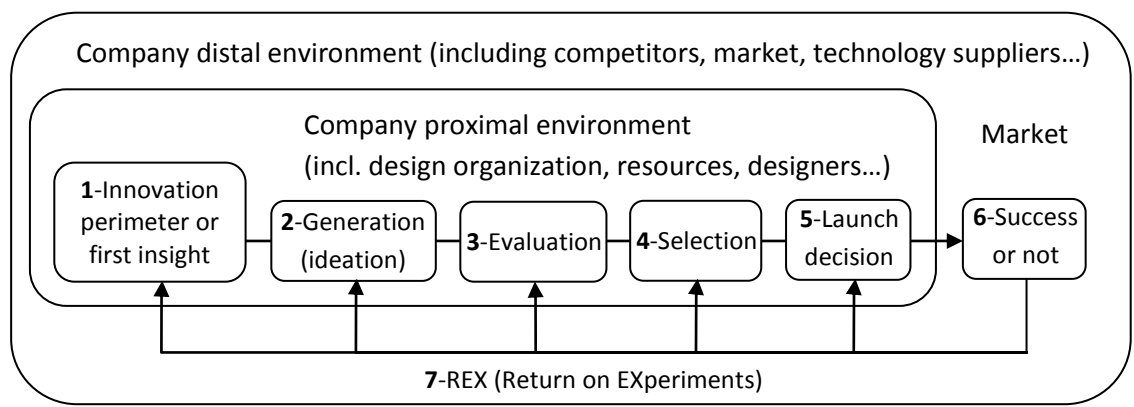

**Figure 1** *A practical company-oriented perspective of design creativity and innovation must encompass 7 research topics*

Working in an Industrial Engineering laboratory, we advocate a more systemic vision of design creativity and innovation in company ecosystems. We present in this paper an attempt to develop and make professional an innovation engineering started 5 years ago in Ecole Centrale Paris. Several research models, experiments, PhD dissertations will be introduced to make innovation value for companies.

We think that research in design creativity and innovation must not only deal with the generation of ideas (or ideation) and evaluation stages of an innovation process (corresponding to stages #2 and #3 of Figure 1). We propose to organize research in design creativity and innovation into 7 research topics organized along an innovation process in company (Figure 1):

- Before idea generation (or ideation) stage #2, one must study which innovation perimeter (stage #1) is worthy to be addressed by the company. This is clearly the domain of innovation or strategic marketing (see for instance (Kim and Mauborgne 2005)) but one should better use their theories and methodologies in design to start with the technological and market areas that are worthy to explore.

- After the generation stage #2, an evaluation stage #3 must be enhanced to be capable to evaluate idea proposals in the company ecosystem: the proximal environment including the company design organization, resources, designers (competences, experiences) and the distal environment (including competitors, market, technology suppliers…)

- Idea selection at stage #4 is another important topic since ideas can come from outside the company (e.g.: crowdsourcing) and the company must know how to assess its potential of value creation in its ecosystem.

- When, how and what to launch on the market (see (Motte *et al.* 2011c)) is another big issue (stage #5) that must not be let to the sole marketers since it involves so much the user acceptance of innovations and the strategical decision to progressively deploy innovations in successive product versions or families to maximize the competitive advantages and the profits.

- What is the success of an innovation in the market and the image return for the company (stage #6) and how to measure them are of the utmost importance for two reasons. First, one must orient our methods in stages 1 to 5 to condition as much as it can be this success. Second, we must also organize continuous improvements of these stages from the learnings of successes and failures.

The consequence of such a company-oriented research on design creativity and innovation (see (Yannou and Petiot 2011)) is that it requires strong industrial connections, design research is then more context dependent, design tools and methodologies are more value-creation requiring in company context. This is why design processes, design knowledge and competences, design organization, design platforms, digital prototyping, design management, and integration of methodologies must be strongly emphasized for that perspective. Rigorous approaches of "action-research" types derived from management sciences (see also (Blessing and Chakrabarti 2009)) are often used to improve design methods and tools in an industrial context.

In the following, we illustrate the aforementioned research topics of Figure 1 by works performed at Ecole Centrale Paris (see in Figure 2 the sections concerned). We start in section 2 (§2) by a recent survey on innovation practice and organizational models led in 28 large companies. The lessons learned about this survey led us to reinforce the belief that there is a need for a new method in agile management of radical innovation projects (section 3). In section 4, we evoke the works on *Usage Coverage Models* (UCM) that allow expressing a space of usage scenarios which is worthwhile to be addressed in the innovation problem setting. In section 5, we propose a methodology of radical innovation in multidisciplinary and company contexts: *Radical Innovation Design*® (RID). In section 6, we evaluate the effectiveness of this RID methodology in performing a large scale evaluation of the project outcomes for the company. In section 7, we derive a selection procedure and organization structure called SAPIGE® to select innovative ideas and projects in an innovation cluster. Finally, in section 8, we evoke a value-driven process and design platform which has been deployed for airplane development projects before concluding.

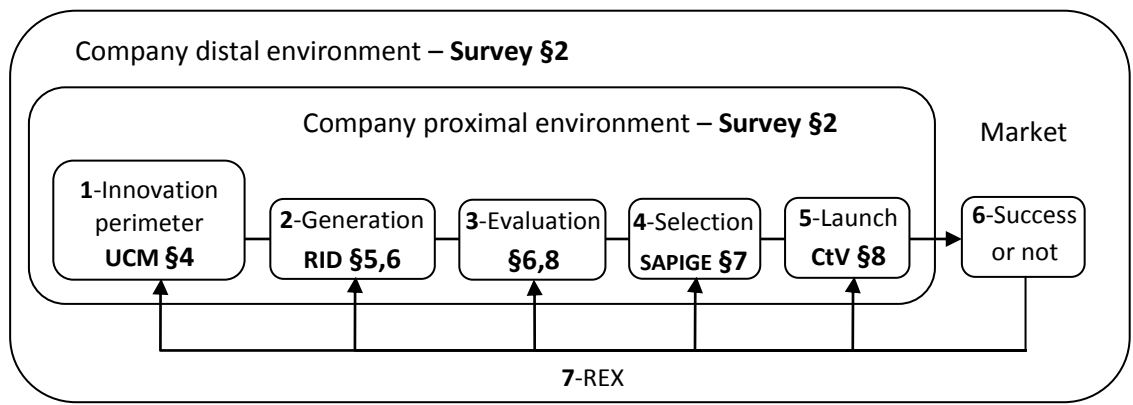

**Figure 2** *Outlook of research works in Ecole Centrale Paris on innovation engineering*

## 2. Survey on innovation in large companies and lessons learned

In a recent survey of 28 large European companies in the industry, we investigated the state of practice in innovation and innovative organizational models in large companies (Cuisinier *et al.* 2012). We interviewed 48 R & D or innovation directors by asking them to self-diagnose their business practices according to the five management areas that contribute to value creation: strategy and business intelligence, organization of R & D, management of innovation processes, innovation culture and management of human resources and R & D, measurement of innovation performance. Our learning are numerous and sometimes surprising. Here are a few thereafter.

Two thirds of respondents reported having profoundly transformed or reorganized their R & D over the past three years. For reasons to support the international expansion, pooling and centralization of research upstream, location in the business unit of applied research and development or reorganization of the development process and resource allocation of R & D. These reorganizations are made in trial and error mode (no apparent method) and reflect a search for greater innovation performance. Indeed, companies face a real difficulty in measuring performance and the benefits of innovation, investment in research that results often lately by the market launch of innovative products and services. Management indicators of innovation or value creation are often of "rear-view mirror" type like the number of patents rather than of "looking ahead" type (able to monitor the value increasing).

In the opinion of the companies themselves, they still fail to know customer expectations and needs. B-to-B companies have a poor understanding of customer needs, their way of thinking, acting and more generally of their habits and lifestyles, as they are focused on the real-time monitoring of consumption. Surprisingly, companies who best know their end customers are suppliers of one or more rows since they must both better anticipate the future demands of their direct customers, and better anticipate the potential and feasibility of technologies in their fields of activities. For these companies, business intelligence and strategic vision result in a continuous updating of business and technology roadmaps.

When asked "who are the actors of innovation in the company?" we can distinguish those involved in the business intelligence fairly well distributed across all business functions (sometimes with a dedicated function to that) and those who choose the strategy that remain largely confined to a top three R & D, General Management, Strategic Marketing and Product Plan. This demonstrates some compartmentalization of the actors that drive business value creation. There is of course the famous use of "open innovation" through collaborations with a number of external collaborators at certain times of the value creation process: academic partners, innovation by suppliers, innovation by customers, innovation clusters, joint ventures, participation in innovation networks, incubators, co-licensing/patent pooling... The rate of uptake of innovation is very diverse, but all the companies want to maintain or increase it. Yet one measures that there is no apparent correlation between the use of open innovation and the innovation performance of the company. In addition, barriers to open innovation, which are well known (sharing of intellectual property, privacy, profit share) do not appear to be the subjects of legal or collaboration innovations to make routine modes of value creation.

Finally, the companies say that the upstream processes of ideas management are poorly organized. Indeed, 47% of companies do not use a methodology for generating ideas. The methods used are the idea boxes and idea contests without, in most cases, budget for the exploitation of good ideas that emerge. Also appalling, the only methods of generating ideas and driving innovation that are sometimes referred by high-level managers are TRIZ and Design to Cost and Objectives.

The companies surveyed acknowledge the fact that the so-called "innovation process" of a company is actually a series of strata or four processes (see Figure 3) with interconnections but acting at different times, with specific strategies, roadmaps and different but interdependent budgets. These four processes are:

1. The process of ideas generating on products, technologies, processes or organization,

2. That of research or technology management

3. That of product lines or programs management or planning

4. The very activity of project management of New Product Development (NPD), which supplies the market with new offers and contributes largely to the creation of business value.

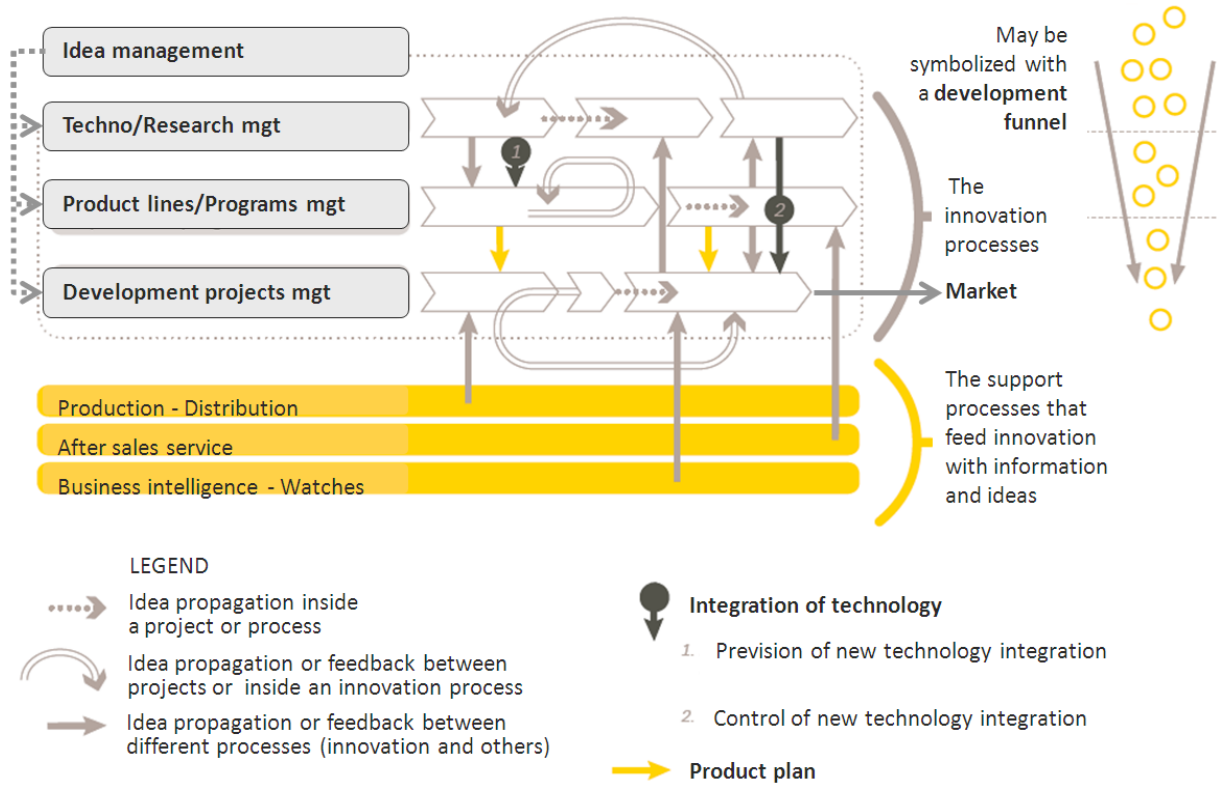

**Figure 3** *Modeling innovation or value creation processes in companies*

It may be noted that a number of business support processes such as marketing, customer relations, after sales service, purchasing and competitive intelligence also contribute more indirectly to the increased degree of effective innovation of the company's products. Note also that according to the ideas (nature, size, maturity), process #1 of Ideas Management can feed processes #2, #3 and #4. All this must necessarily be organized within the company with the collection process of ideas and transfer of ideas within mature roadmap that invest and plan their development and their deployment in research, product lines or project development. However, these transfer processes are currently poorly organized and coordinated so that the process of generating and collecting ideas as we have seen. But there are organized and standardized processes within the company such as product development. These are step by step processes with intermediate outcomes expected, so-called "stage and gate". It turns out that these processes are as well necessary to ensure a minimum quality and coordinate development activity on a complex project, as sometimes too rigid and not very permeable to new ideas and opportunities that would upset too much a strategic positioning or that would appear during the project. In our book (Cuisinier *et al.* 2012) we have thus established a model of five funnels types of innovation (such as that shown in Figure 3) who have their own characteristics, rationale, but sometimes with different benefits. We have baptized them: the funnel "standard, but very fertile", the funnel "standardized with high attrition", the "heroes" funnel, the funnel "mixed (or dual device)", the funnel "agile and permeable." We have shown in (Bertoluci *et al.* 2013) that these different forms of funnel are part of a natural and logical process of cyclical development in five successive stages.

It has been concluded by companies that the ultimate goal is to be able to organize an agile management of innovation (or value creation). Our conclusions of this survey (Cuisinier *et al.* 2012) are that it is now time, in the years 2010-2020 to *orchestrate and professionalize a supply chain of innovation*. Concretely, this means to:

- Invest to script the future, i.e. to develop at several levels of the company strategical visions and prospective scenarios – roadmaps – so as to "stand by and be one step ahead". This requires the implementation and the sharing of a systemic vision of the company.

- Broaden the definition of the innovation strategy to the different business functions.

- Develop forward-looking indicators for measuring innovation performance.

- Systematically exploring new territories. This requires addressing several locks:

    o  To make systematic, in early phases of innovation, the confrontation of the *possible* (emerging technologies, research), the *feasible* (what is achievable with acceptable cost / know-how / time by the company in its ecosystem) and *desirable* (knowledge of the market needs or trends before competitors). In other words, it needs to adopt and generalize the principle of dynamic control or exploration / exploitation of innovations and value leads.

    o  To develop skills of "innovators" between Marketing and R & D.

    o  To think of innovation in terms of innovative organization and maximize the potential of each employee in orchestrating innovation through decompartmentalization of functions.

- Embody a culture of innovation by the resources allocated and operational involvement of senior management.

### 3. A need for a new method in agile management of radical innovation projects

Having supervised for years innovation projects with a mix of people - engineering students, students of business schools and industrial design -, and following the same observations than industry managers themselves about their value creation processes, we were led to ask ourselves about the knowledge and methods to teach our engineers faced with the challenges of tomorrow's innovation.

We made the observation of a significant compartmentalization of disciplines interested in innovation and value creation for companies. In companies, this compartmentalization also exists in the sequential management of activities: planning of research projects, planning of development projects (called product planning), management of the conceptual design phase where 80% of the potential value creation is committed, management of product-process detail design, without omitting the sporadic interventions (at least on engineering projects) of the most creative people like industrial designers. Given the importance of the challenges engineers will face in the near future, especially with regard to both societal and system innovations, we believe that a high-level engineer needs to know how to properly set and solve a situation radical innovation in the context of a given company. We must teach them the knowledge and expertise of business innovation at the interface of disciplines and business processes.

We started by analyzing the limitations of current classical engineering design approaches and those of product development from management science, and the limitations of their juxtaposition in a company (which is often the current rule) in the particular case of radical innovation (see publications (Motte *et al.* 2011a, Motte *et al.* 2011b)). Indeed, in the case of radical innovation, the transmission of a marketing brief originating from a product-service planning to an engineering department, that starts at this time the development project, is not at all satisfactory. We show in (Motte *et al.* 2011c) that planning comes up here to introduce a new function on the market or to integrate a mature enough technology exiting an internal research project to result in a brief that provides a strict direction of the innovation perimeter to explore. In short, the part of innovation that remains to be done in the conceptual design (which is generally attributed to engineers) is often minor compared to the strategic choices of prior planning. In addition, there is no evidence before the project begins that this brief is feasible (proof of concept) and produces an experiential value for users (proof of usage and esteem value). This planning is, in short, a perimeter frame in which to innovate, made to reassure companies that wish to minimize the risk of launching a new product or service. We believe that there are other development or launching strategies to minimize these risks (see (Motte *et al.* 2011c)) but that it would require to integrate and merge the two stages of product planning and conceptual design in a radical innovation project to allow two-way influence between the opportunities for value creation (mainly carried by marketing but not only...) and feasibility (carried by design) and perceptions and user experience (brought by the industrial design) that emerge from design choices during the development project. The scope within which it would be legitimate to innovate (that we further refer to *perimeter of ambition*) should be determined in a more opportunistic way during investigation of markets and existing usages and also in the light of emerging adequate conceptual design solutions. We believe that multidisciplinary teams (marketers, design engineers, industrial designers) must do this work together in a mode of agile management. We show in (Motte *et al.* 2011c) that this requires four conditions:

1) A dynamic strategical alignment of company over brief opportunities emerging during the project, and which can change the values, the market or the traditional partners of the firm,

2) A reorganization of the company especially in terms of processes and tasks of marketing and design,

3) A change of the organization that must overcome some obstacles to work habits and personal influence stakes. These aspects are well known in companies that have decided to be new entrants or pioneers in innovation, shifting the traditional logic of competition. The Blue Ocean Strategy (Blue Ocean Strategy) here gives a very clear framework to bring an organization to produce radical innovations.

4) A new mode of innovation in multi-disciplinary team to truly co-innovate by participating in a synergistic way to defining the perimeter of ambition of the brief and of the conceptual design rather than to innovate by business silos in juxtaposing the solutions of every participating designer. This means, as we can see further, a new mode of interaction between businesses to speak a common language for project reviews and make more collective decisions.

## 4. How to target actual needs to drive creativity insights? Design by Usage Coverage simulations

The marketing literature has advocated for decades that new products should be designed for intended and anticipated consumer usages. Dickson in 1982 (Dickson 1982) pleaded for a renewal of marketing research for better segmentation by considering usage situations: "*A recent comprehensive state of the art review of market segmentation concluded that the field has become too fixed in its ways and that new conceptualizations of the segmentation problem should be explored. One convention that bears examination is the equating of market segmentation with customer segmentation. Markets can also be subdivided by usage situation. Although almost every conceivable person-based characteristic has been used to segment markets over the last decades, there has been a disturbing lack of consideration of the usage situation as a basis for defining product markets and modeling consumer choice behavior.*" Despite the fact that the value of considering usage in marketing and engineering studies has been noted in the literature, little has been done to merge integrated approaches for resultant operational design methods. Indeed, when one seeks to design an adapted product, product-service or product family for a market, two families of methods are available. The first method is *design optimization* of intrinsic performances and the second is building a *prediction model of the market share* after conducting a tedious market study. They both they suffer from a lack of realism in terms of simulation of personal usage needs. Optimization is mostly based on averaged expected performances; it is also independent of specific users' skills and does not try to model neither sets of anticipated usage scenarios nor competing products on the market. Additionally, marketing choice and market share models require tedious market investigations assuming an existing market experience of products, which is not the case for disruptive innovative products. In contrast to tedious market studies which assume an existing market experience for products, and optimization approaches based upon static product performances, we propose an adaptable approach to a market to designing a product, product-service or product family. This is why a usage-centered model-based approach, as proposed in Figure 4 (see also (Yannou *et al.* 2012a)), has value in an innovation process to determine an innovation perimeter which is worthy to explore before the idea generation and selection stages (see stages 1,2 and 3 of Figure 1). We name this new approach "*design by usage coverage simulation*" based on a *Usage Coverage Model* (UCM). The principles and ontology of UCM has been set in (He *et al.* 2012, Yannou *et al.* 2009). Next, the principles of "*design by usage coverage simulation*" has been set in (Yannou *et al.* 2010b). First, one generates a usage scenario space for a set of representative users. Our approach does not *a priori* assume technical attributes when building the space of usage scenarios. Consequently, it is then more likely to compute solution-independent market models and to serve as a decision aid in case of innovative designs. Next, considering a set of candidate products, possibly of a scale-based product family, one proceeds to make set-based computations of feasible usage scenarios, provided that physics-based simulations of performances are possible. The comparison between the expected and feasible usage scenarios at the scale of a single user, considering the level of delivered performances and product price, leads to *Usage Coverage Indicators* (UCIs) and finally to a preferred product best covering the personal usage scenario space. UCIs have been proposed in (Wang 2012, Wang *et al.* 2012) as a way to measure the potential to satisfy the entirety of anticipated usage scenarios, for a sole user or for multiple users, for a sole product or for a scale-based product family. A definitive advantage of our approach is that the personal usage coverage simulation of a customer/user depends upon his/her profile, notably skill abilities and usage contexts. User profile is almost never considered in performance models in design engineering research but which can dramatically influence performance in real life situations. The objective is to simulate how people evaluate if a given product is capable of covering the entirety or a sufficient subset of the usage contexts he or she is able to anticipate.

At the level of a targeted consumer group, the approach leads to a market share simulation of competing products or members of a scale-based product family. It is then used as an evaluation tool to confirm the proof of value of the chosen conceptual design; it corresponds to stage 3, 4 and 5 of Figure 1. Our model-based approach has been thoroughly illustrated by the usage coverage simulations for the design of a jigsaw, for a sole user and for multiple users, for a sole product and for a scale-based product family (Wang 2012, Wang *et al.* 2012).

Used as a simulation tool to define a subset of usage scenarios that are badly covered by competitive offers, the UCM approach aids at identifying a *perimeter of ambition* which guarantees an existing need and a profitability (*proof of value*). Then, it perfectly corresponds to stage #1 indicated in Figure 1.

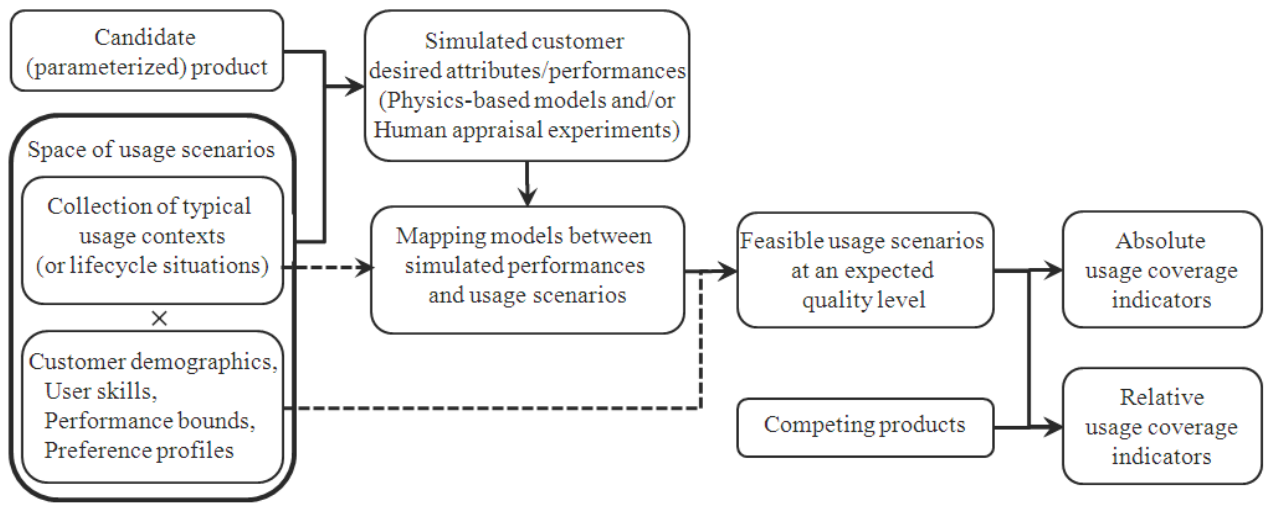

**Figure 4.** *Design by usage coverage simulation framework*

## 5. Proposal of a methodology of radical innovation in multidisciplinary and company contexts : Radical Innovation Design

Shah et al (Shah *et al.* 2000) note that "*A wide range of formal methods have been devised and used for idea generation in conceptual design. Experimental evidence is needed to support claims regarding the effectiveness of these methods in promoting idea generation in engineering design.*"

After years of design education in innovation, we have defined a methodology called *Radical Innovation Design* (RID). This methodology for innovating products, services and/or business models in industrial contexts, has been taught for 5 years at Ecole Centrale Paris, a Master level engineering school in France. It focuses mainly on the design stages from the initial need statement to the choice of the conceptual design solution, which is the resulting design outcome that is supposed to maximize value creation within a specified industrial context.

In addition to a number of design principles that already exist in the literature, RID is based on a set of new principles and tools, all of them organized within a design process of radical investigation of the highest value creating design solutions. We started from the statement that current design methodologies have a number of shortcomings when addressing a practical innovative product or service development process within a given company (Motte *et al.* 2011a, Motte *et al.* 2011b). Furthermore, the result of the conceptual design stage is generally much more the result of a random or poorly organized but highly creative process than of a radical exploration of value creation opportunities within the company context (see (Motte *et al.* 2011c)). On the other hand, idea generation methods have been quite extensively studied but, most of the time, the quality of design outcomes is assessed without consideration of company context. Indeed, the quality criteria adopted sometimes concern more the means (number of generated ideas, intensity of conceptual design process) than the quality of the chosen conceptual design solution itself. In addition, previous research does not point out clear levers to improve the design process or explain the reasons for the greater or lesser value/quality of innovative project outcome(s). Finally, output quality cannot be an objective assessment, as it is essentially a perception of expert jury members or steering committee members; this subjectivity in outcome value perception has not often been taken into account.

Radical Innovation Design (RID) methodology can be used when the company objective is to *innovate fundamentally*. This requires to take into account the company's positioning in an ecosystem, i.e. it has a strategy, a market presence and a brand reputation, an existing product-service-technology portfolio, competitors and suppliers, and it disposes of certain financial, industrial and intellectual assets (including innovation know-how, technical competences, patent portfolio). With regard to this concrete company context (almost never considered as an input in an innovation process), how is it possible to innovate as much as possible, creating positive differentiation in the market and changing the conventional rules of competition? RID, it should be noted here, is fully compatible with the principles of Blue Ocean Strategy (BOS) marketing strategy developed by Kim and Mauborgne (Kim and Mauborgne 2005).

The *radical* nature of the RID methodology may be understood in terms of a systematic exploration/exploitation process in the following four stages:

1) Exploration of value creation opportunities around the *initial idea or statement* (techniques issued from both strategic marketing and economic intelligence approaches). The *initial idea/statement* is systematically redefined in a more legitimate *ideal need*. Within this new exploration perimeter, existing usages, needs and product experiences are populated, investigated and benchmarked so as to yield stage #2.

2) Definition of a promising and coherent *perimeter of ambition* which is a subset of the aforementioned *ideal need*. This *perimeter of ambition* must represent an opportunistic potential of value creation in the context of the company ecosystem.

3) Definition of some value promising product-service scenarios, also called *briefs*, starting from the *perimeter of ambition*. These briefs must be qualified (often by storyboards) and quantified (size and willingness-to-pay of the markets) using at best the previous investigations performed.

4) For each brief studied, a systematic listing of *value tracks* and *value drivers* further called *innovation leads* (see the use of RID methodology in the context of EADS company (Rianantsoa *et al.* 2010a, Rianantsoa *et al.* 2010b)) are performed. Each *innovation lead* is in turn investigated in the form of a systematic creativity workshop. Further, findings are combined into consistent design concepts which are, in turn, sketched or prototyped and assessed in their whole.

Finally, the RID methodology is organized, following Herbert Simon's approach, around a two-part macro-process: the *problem setting* macro-stage and the *problem solving* macro-stage. Figure 5 represents these two macro-stages in the so-called *RID innovation wheel*. For the four radical exploration/exploitation stages previously evoked, the two first belong to the *problem setting* macro-stage where no evocation of solution is tolerated, and the two last belong to the *problem solving* macro-stage. The RID innovation wheel spans from the *initial idea* or statement to the *feasibility and innovation dossier* passing through intermediate results such as *ideal need, perimeter of ambition, brief(s), concepts*. In practice, a series of micro-stages are defined and documented with expected intermediate results and reports, practical examples for inspiration and a toolbox.

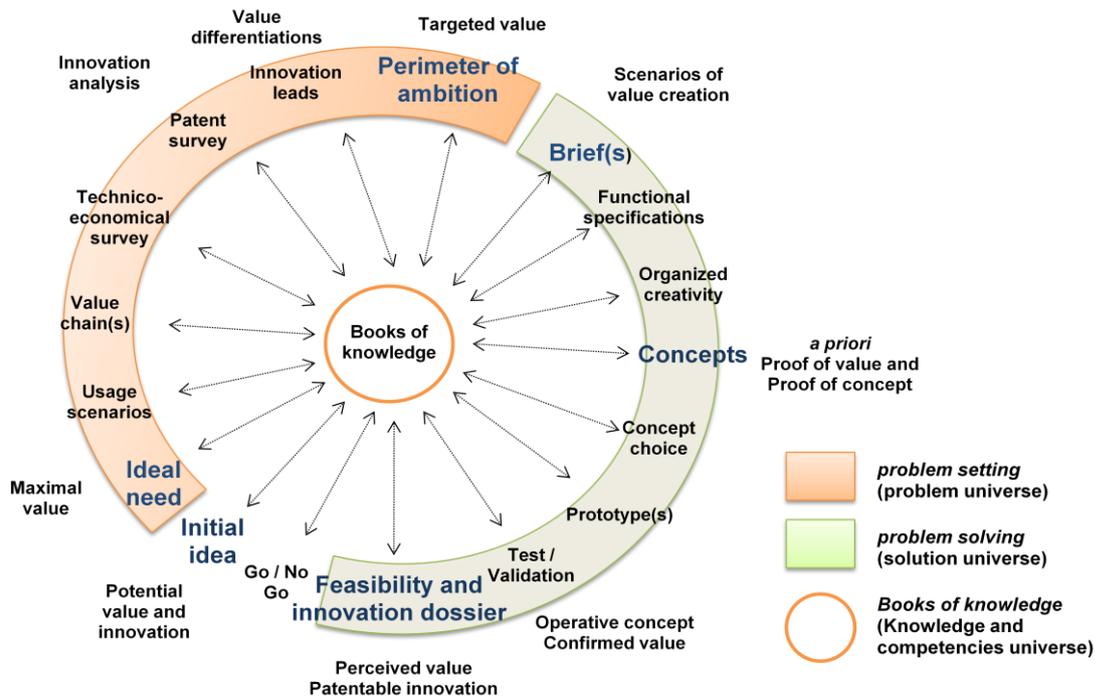

**Figure 5:** *The RID innovation wheel: From initial idea to feasibility and innovation dossier… through ideal need, perimeter of ambition, brief(s), concepts*

A determining concept of RID is to consider the conceptual design stage as an *investigation process*. Investigation is understood as exploring all potential leads and refining and evaluating conceptual designs as long as they appear to be potential value makers. This investigation and conceptual refinement process is known as *issue-based design* or *question-based design* and has been well developed by Bracewell, Aurisicchio and Wallace in the Dred system (Aurisicchio and Bracewell 2009, Aurisicchio *et al.* 2008, Bracewell *et al.* 2009) or in CK-theory by Hatchuel and Weil (Hatchuel and Weil 2003). But RID has practically developed this investigation spirit with the constant building and reinforcement of proofs. Three types of proofs are defined:

- The *proofs of concept* for bringing evidence that "it works or it is likely to work in situations the service is expected to be delivered",

- The *proofs of value* for bringing evidence that "it is differentiating from the existing solutions in terms of service utility as well as new satisfied needs, on large and creditworthy market segments",

- The *proofs of innovation* for bringing evidence that "the invention may be protected and the innovation may be communicated, perceived, understood and valorised, i.e. it corresponds to a certain willingness-to-pay.

Of course, the *proofs of concept* can only be defined as soon as solutions start to emerge within the problem solving stage. But *proofs of value* and *proofs of innovation* can be thought of as soon as the ideal need perimeter (maximal value) is defined and, further, a perimeter of ambition (targeted value). This maturity evolutions of proofs of value, innovation and concept are represented around the innovation wheel in Figure 5. Following this idea, we have recently proposed a series of performance indicators to practically measure these three proofs (see (Zimmer *et al.* 2012) and section 7) at the stage of idea and project selection in an innovation cluster, and also for the coaching of the selected and financed project so as to reinforce the levels of proofs and prepare a solid business plan. It is then become a real design principle: to reinforce the level of these three proofs to result in a design concept which is likely to be implemented by the company and to become successful on the market.

A corollary concept of this *investigation* process is the necessary *documentation* and *knowledge management* (including the competences of the design team members) and the constant evaluation of the probability of coming up with a conceptually useful design. This documentation can be supported by an issue-based information system like the Dred platform. But we also refer to the Intermediary Design Objects (IDO) concept first proposed by Jeantet and Boujut (Boujut and Blanco 2003, Jeantet 1998) which enables the designers to be influenced in their choices between different concepts (they may be physical or virtual prototypes, sketches, questionnaires, etc). The quality and pertinence of these IDOs determines the quality of the design outcome(s). CK-theory (Hatchuel and Weil 2003) also proposes strategic views of managing design knowledge and competences. Finally, Thompson and Paredis (Thompson and Paredis 2010) propose a relevant Rational Design Theory (RDT) which consists, as RID, in maximizing the expected value of utility of a design concept. But very few theories or frameworks exist that propose performance indicators to guide the *idea generation*, *evaluation* and *selection* (see stages 2, 3 and 4 of Figure 1).

Several other RID concepts (if not principles) need to be mentioned briefly here:

- *Usage* is the first space to navigate in before *functions* which, at the initial stage of innovation, over-constrain the design space. The RID *problem setting* macro-stage is of course fully compatible with the *Usage Coverage Models* (UCM) presented in previous section, to better locate sets of usage that are worth being covered by the design solution (He *et al.* 2012, Wang *et al.* 2012, Yannou *et al.* 2009, Yannou *et al.* 2010a).

- *Books of design knowledge* (inventories or diaries of *Intermediary Design Objects*) must be generated when possible, because they are also a value creation within an innovative design project (e.g. books of patents, books of technologies, books of concepts...).

- A new method of *design collaboration* is encouraged to avoid *silo innovations* in the different juxtapositions of novelties and disciplines concerned. We advocate that design team members and disciplines share their conceptual pathways map in a *co-innovation process,* thus sharing important decisions and trade-offs and fostering innovation at a higher architectural level..

Considering the aforementioned RID principles, Figure 6 summarizes all the value results that may be expected at the end of an innovative design project. This transformation between input data (the initial idea and the company and project ecosystem) and project value results is called *RID value machine*.

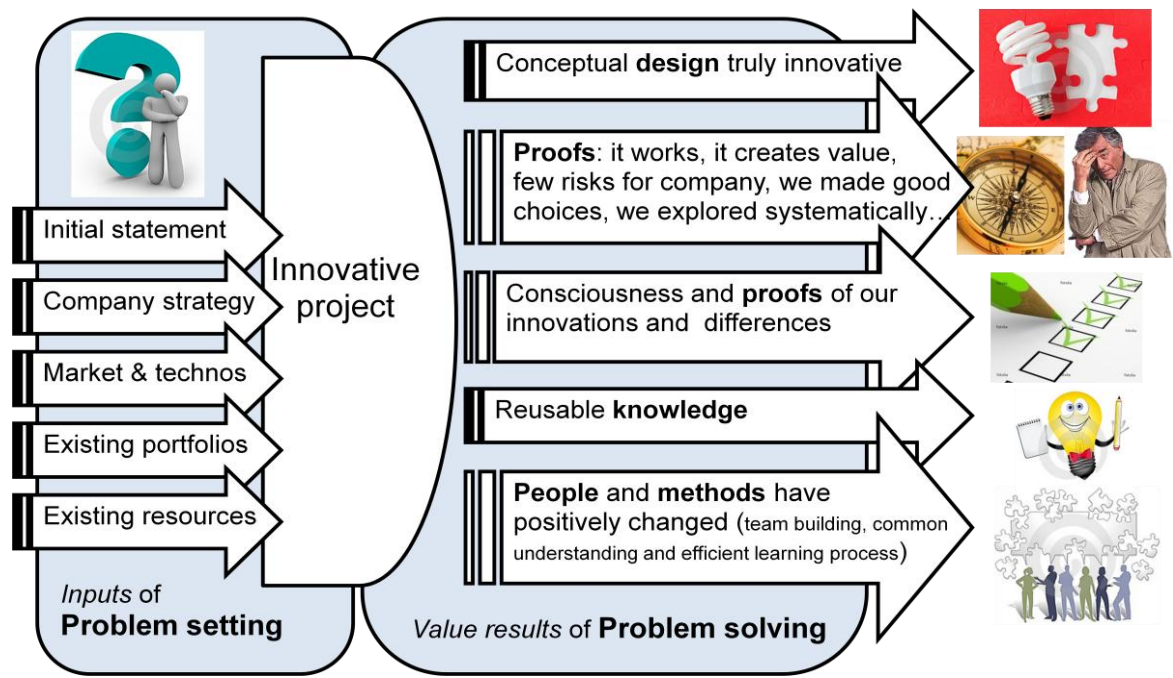

**Figure 6:** *The value machine of an innovative project using RID methodology*

### 6. How to measure value created for companies in an innovation project?

Following our principles of design research methodology (Blessing and Chakrabarti 2009, Yannou and Petiot 2011)), search for methodology effectiveness, and search for proofs of value, concept and innovation for company, we have tried to renew somewhat the research protocols on innovation management and idea generation in (Yannou *et al.* 2011, Yannou *et al.* 2012b). Our observation may appear brutal. Most of the existing literature in design engineering considers the sole creativity management or idea generation, often starting from a well stated customer need and design issue. In real life, companies face *fuzzy front end* situations (see (Motte *et al.* 2011a, Motte *et al.* 2011b, Motte *et al.* 2011c)) where the design problem remains to be set in terms of need justification and market profitability. Secondly, maximizing quantity, quality (often not well defined) and diversity of ideas outcoming creativity workshops is far from being correlated with a successful product development and market launch. This is why we propose to include in studies on innovation management the

**15**

problem setting stage to identify need and market opportunities as well as the result of ideation, conceptual design, evaluation and selection stages. Then, we propose to evaluate the single innovative design that has finally been chosen – whatever the quantity and diversity of generated ideas -. Moreover, one extends the value assessment to the innovative project outcome and this value assessment is made in the context of a given company ecosystem (see Figure 6).

The objective was then to experimentally assess the effective value creation of the resulting selected design concept in the context of an expected radical innovation in a company ecosystem. We have proceeded to large scale experimentation on possible drivers – endogenous to the project - of a successful radical innovation for a company. We had to define a given innovation management framework to make things comparable in terms of identified stages and intermediate results: it is the *Radical Innovation Design*® methodology framework exposed in section 5.

Usually, in any research work of that kind the designers involved in the innovation process are loosely defined in terms of their initial training, knowledge and skill about innovative design and motivation on the project. In addition, most of the papers existing in the creativity measurement and engineering ideation usually poorly define the design stages and tools which have been used with a given intensity so as to characterize the design outcomes. However, both aspects are truly influential on the quality and relevance of the design activity. In our work (Yannou *et al.* 2011, Yannou *et al.* 2012b), all the design teams have been taught with Radical Innovation Design stage and gate process and corresponding tools. But, constrained by the tight project duration, they were not compelled to strictly follow each stage, and to deliver an expected result at each gate. In addition, we have proposed not to follow the traditional threefold quantity-quality-diversity nature of the ideas generated for the sole ideation process. We have rather proposed an enriched "model of value creation in company context" to directly assess the final chosen innovative design concept, including generation, evaluation and selection processes.

Providing these two points – the taught but not imposed RID framework or process and the value model of the preferred innovative design concept – the experiments referred herein amount to assess whether design methods and gate deliverables recommended in RID (called *means* of the *problem setting* and the *problem solving* processes) effectively influence and favour the effectively delivered value of the selected design concept (which is a part of the *results*). This value is in fact evaluated by expert *jury members* – being for half of them executives of companies - in a limited time having in mind the expected value creation in the company ecosystem, and not only the sole apparent innovativeness of the idea or concept in itself. An important note is that the evaluation of *results* (the value created for the company) has been made independently of the *means* employed (i.e., the more or less observance of the RID prescriptive methodology) by both faculty and company R&D managers. An additional objective is to discover key factors and causalities between phenomena – design acts and value results - about innovation management *in the context* of *design team* features, *project type* features and the degree of *assimilation of innovation* principles and tools by the project participants (here, assimilation of RID methodology). Our observation protocol of innovative projects is summarized in Figure 7.

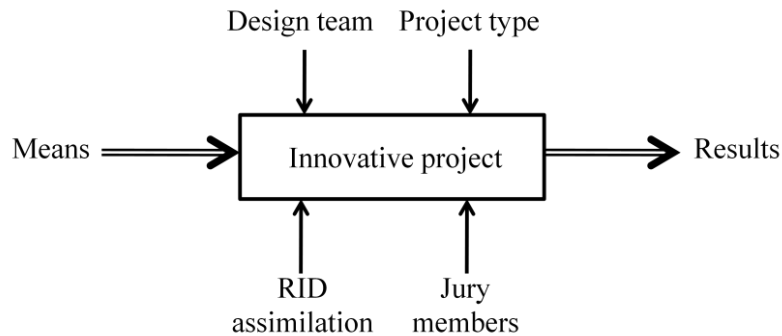

**Figure 7.** *Our observation protocol of innovative projects*

For this purpose, a test bed of 19 design projects of 5 different types, after a common RID methodology training, with 86 students has been performed (see (Yannou *et al.* 2011, Yannou *et al.* 2012b)). Sixty-one variables were screened to characterize the project *means*, project *results*, and to detect conditional probabilistic dependencies with *design team* features, *project type* features, *innovation principles and tool assimilation* levels, and *jury member* features. These observations have been summarized into different Bayesian Networks (BNs) through a primary unsupervised learning process and additional finer supervised learning processes. These BN models provided many causal validations and sometimes revealed unforeseen links between design means, design results (effective produced value), contextually dependent on the project type, the design team composition and the jury themselves. This non-trivial material, presented briefly here, is an approach we recommend when exploring the production of innovative value in industrial contexts. Indeed, the importance of context cannot be understated, as a variety of variables influence the innovation process, often with conditional or non-linear effects, which justifies *a posteriori* the choice of Bayesian Networks. Next, we have proposed to study the conditional dependencies between these variable modalities without any preconception. Intensive Bayesian Networks learning processes and further "what if" Bayesian inferences have been performed.

Our results have been rich and they may appeal to revisit the results found in corresponding scientific literature because the traditional experimental protocols appear now being far too simple.

The first finding is that for radically innovating the quality of the problem setting stage is determining and especially (see Figure 8) a *proper definition of ideal needs* – what people fundamentally would need or expect -, an *extensive knowledge of usage practices* in the neighborhoods of the *initial idea*, a good *identification of stakeholders* of existing product-service value chains and a good *definition of present usage contexts*. This finding somewhat contradicts the Design Thinking attitude of "*do it as soon as possible and improve it if necessary*" to a more business intelligence attitude of "*explore the available information before to dig in one precise direction*".

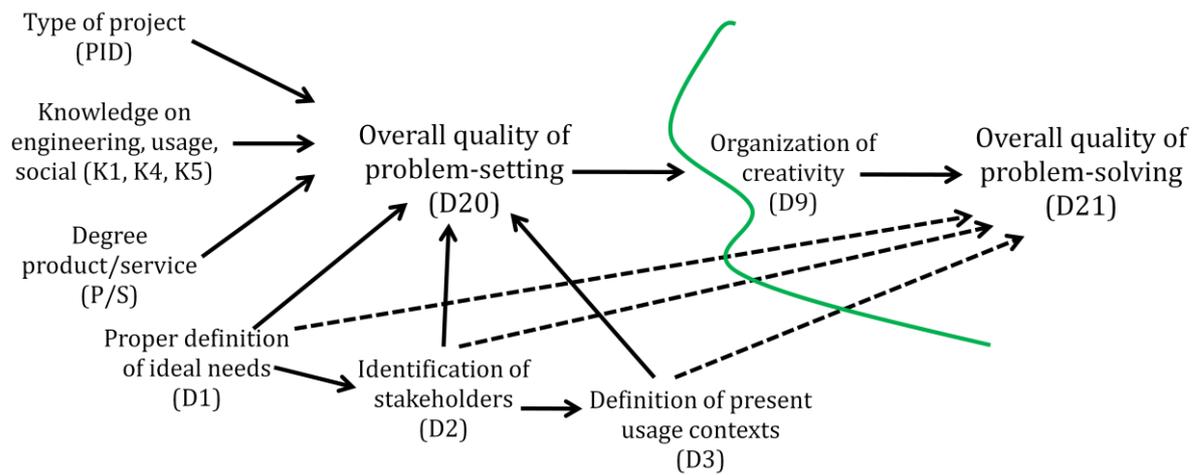

**Figure 8.** *A major finding of study: the overall quality of problem setting process dramatically influences the overall quality of problem solving process*

A second major finding is that the importance of design context, including the company ecosystem, the type of subject – product versus service, unusual knowledge for the company -, the team skills and motivations - cannot be understated. We show that such variables influence the innovation process and modulate effects, often with conditional or non-linear effects, which justify a posteriori the choice of Bayesian Networks for learning and simulation. To give an example, a good level in the design team of *engineering knowledge* and *usage knowledge* are major success factors for certain types of need-driven radical innovations (the technology-push and market-pull ones). For these expected innovations, one should particular take care of that, rather than expecting too much from creativity workshops. It also means that an innovation process cannot be driven by a rigid prescriptive methodology imposing a fixed stage-and-gate process. This is why we believe now more in an agile innovation process that must be consciously adapted beforehand after an analysis of the innovation stakes of a particular project. A major issue we will now focus on.

Lastly, the third finding of our study is the necessity to beware of idea quality or innovative project assessments. We realized that the jury members have been abusively impressed by certain projects that did not deserve it. Few existing papers really consider the experience of people performing innovation appraisals. In fact, the traditional fast design reviews to make decisions at the end of an innovation project – or a conceptual design stage – could be counter-productive since several rebound-effects can explain that flashy presentations of a chosen design concept of medium quality can be preferred to more apparently dull justifications of a good problem setting and an exhaustive brief and concept exploration – good problem solving -. However, our experiments clearly show that a good problem solving process depends on a good problem setting process and that these design process assessments must be done carefully in screening appropriate design documents. This consideration on how to proceed for a design evaluation to make the decision of developing or not a chosen design concept is also of major importance. We clearly conclude here that it is an issue to further study and that a shallow design review of experts may be misleading.

## 7. How to select promising innovative ideas and projects in an innovation cluster?

A first industrial extension of Radical Innovation Design® methodology has been brought in the context of an innovation cluster of 70 institutions dedicated to promote innovation in products and services for elderly people (Zimmer 2012). An innovation cluster is a kind of open innovation organization aiming at sharing knowledge, networking and promoting a more efficient multidisciplinary innovation. Here, the main mission of the cluster is to select the most promising product-service innovations in the contexts of an existing company or the creation of a start-up company. On the principles of *Radical Innovation Design®* methodology, it was proposed a system, called SAPIGE®, for selecting innovation projects that present sufficient *proofs of value, innovation and concept*. Precise definitions of these categories of proofs were given (see (Zimmer *et al.* 2012)) and a list of 22 *evidence* selection criteria has been accurately defined. This list of criteria and health, economic, technological and social values that underlie them were carefully recorded in the appendix of the founding charter of the association of the cluster. Indeed, these texts have led to the emergence and to share an ethical and economical vocabulary and positioning for all the cluster stakeholders composed of many different cultures and businesses (medical doctors, robotics researchers, large companies, SMEs, associations of users and retirement homes, social security institution, private and public investors in innovation ...). These actors were finely categorized and ultimately distributed according to three commissions of experts for participation in innovation juries: users, innovation, industry. In practice, this selection list is used as a two-filter procedure (see Figure 9) which includes the two macro-stages of the RID innovation wheel, namely:

- The *problem setting* on the one hand, where only the quality of the problem setting of innovation is important in a first time. The *usage value* and *innovation value* of the idea and of its *perimeter of ambition* are assessed independently of the progress of the implementation of a solution.

- The *problem solving*, on the other hand, where *proof of concept* and *proof of profitability* (a part of *proof of value*) are evaluated from the solution defined or sketched.

Different jury commissions of experts are composed depending on the nature of the filter in the selection procedure (see Figure 9). Three benefits are expected from the innovation jury: a label (benefit of scientific credibility, and therefore a positive image), a non-refundable financial contribution and finally a one-year coaching from a professional in innovation management to improve the potential of success of the project. The three benefits can be combined in whole or part. An algorithm based on the accumulated value of the proof categories during the two stages of selection leads to the possible gratifications. For example, a project with an excellent problem setting is excellent but little or no problem solving will be rewarded with financial assistance as well as a one-year coaching to advise the project carrier at best and maximize the development of the project. As such, the selection list also serves as a diagnosis tool on the evidences to strengthen and, thereby, serves as a support check-list during the one-year coaching. During this year, the cluster having a structured service offering indexed by contributions to evidence reinforcement, the innovation coach of the cluster may suggest ways of improvement by activation of certain services. For example, a cluster member which is a consulting firm in patents may assist in the identification and protection of innovations, or a geriatric hospital or an association of nursing homes can provide testing grounds for new technologies.

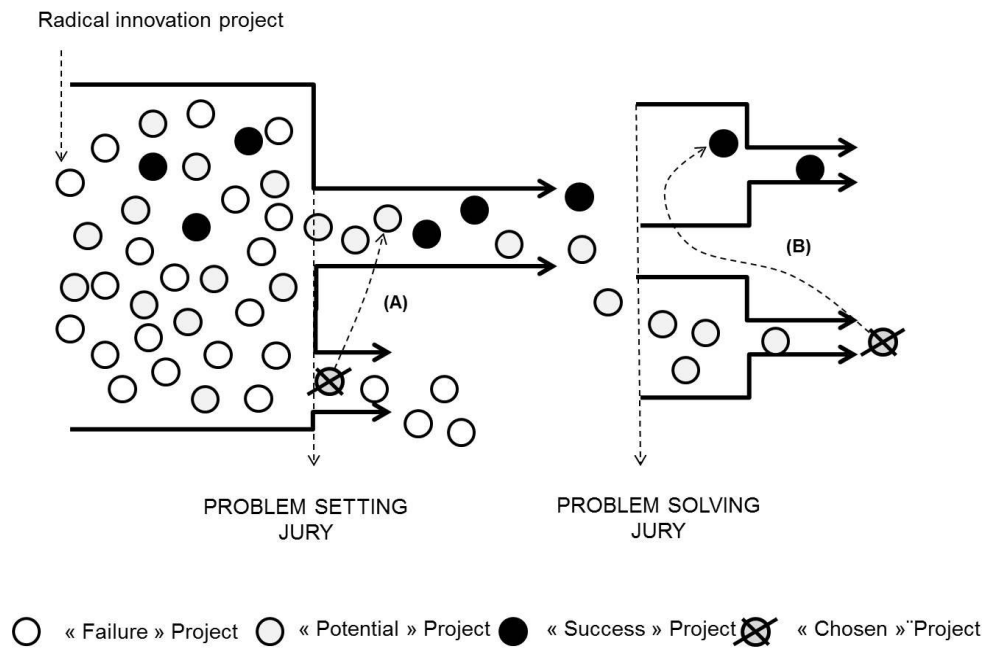

**Figure 9.** *Radical innovation project selection procedure*

Our selection model is made up of a procedure and a chart. It offers three advantages. First of all, it makes it possible to more effectively finance projects with high potential for success in the market based on the evaluation of proofs of value, innovation, and concept. Secondly, the chart that we propose has shown that it gives a clear frame of reference in order that the experts from the juries might develop a more collective vision of the expectations of a radical innovation project. Finally, the results of our satisfaction questionnaire, individually administered to the experts after their testing of our model, show that the procedure we are proposing is effective. First, it gives structure to a discussion on the interest of allotting funding and/or support to an innovation project. Second, the use of an evaluation chart allows the experts to create a common language in order to measure the success of a radical innovation in the market.

## 8. How to deploy a value-driven process in an airplane development project?

A second industrial extension of Radical Innovation Design® methodology has been brought in the context of airplane development projects by a value-driven process following RID principles named *Concept-to-Value* (CtV) (see (Rianantsoa 2012, Rianantsoa *et al.* 2011a, Rianantsoa *et al.* 2011b)). It illustrates the *evaluation* and *decision to launch* research topics in Figure 2 (stages 3 and 5).

As it has been already highlighted in sections 2 and 3, the integration of product planning and conceptual design stages is a goal to reach so as to explore quickly more innovation opportunities. This is the objective of CtV methodology to support integration of product planning and conceptual design stages. Indeed, the objectives of the designers in a commercial aircraft development project like Airbus have above all been the achievement of the aircraft mission and the certification rules. Today, the competition between airplane manufacturers leads to bring more added values to the stakeholders. Other types of values have then to be considered as higher level objectives like the ground operations and maintenance costs, the environmental impact, the image, the security and the autonomy. Therefore, the conceptual design must be driven in the perspective of value creation objectives to define the perimeter and organize innovation from the first airplane specification sheet to a satisfactory dimensioned architecture. Consequently, the traceability of value contributions of design concepts to the entire airplane value must be better supported. An explicit enriched representation of the value model and the targeted stakeholders is then built. A strategical alignment transforms value targets into marketing business strategy and low level innovation strategies that drive design concepts development.

Our *Concept-to-Value methodology* (CtV) is based on a generic model representing, on the one hand, *the multidisciplinary knowledge, the problems and the solutions,* and, on the other hand, the *potential values* they generate for the stakeholders. This model describes the different *Intermediate Design Objects* that are generated both in *product planning* and *conceptual design* stages and must be integrated and systematically assessed in terms of potential of value creation. This model is named PSK-Value (for « *Problem* », « *Solution* » and « *Knowledge* »); it allows to define (see Figure 10):

- A common language for describing business and engineering elements that are generated in the preliminary stages. The *PSK-Value model* (see Figure 10) is used to represent and specify items from *Business teams* (i.e., Marketing, Product Strategy, Program Management...) and *Engineering teams* (Design Engineering, R & Ts, Manufacturing Engineering ...) of new projects. These elements describe the *Knowledge*, *Problems* and *Solutions* explored.
- A systematic approach to measuring and mapping the value and maturity of the multidisciplinary elements.
- A common global process of convergence management and of the collective project progress.

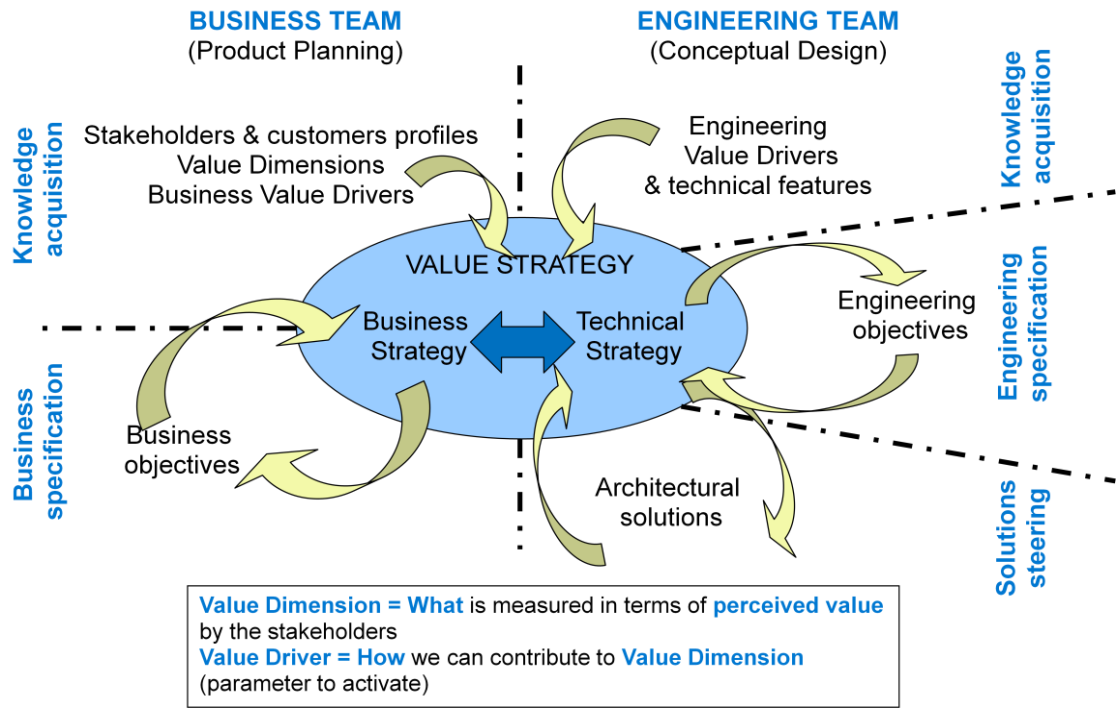

**Figure 10** Ontology of the PSK-Value model (Rianantsoa *et al.* 2011a, Rianantsoa *et al.* 2011b)

Gradually, in different phases and iterations, the Business and Engineering mutually increase their knowledge by sharing and integrating it (exploration of new engineering parameters or technical solutions from business value drivers and vice versa). They increase their degree of convergence by establishing common strategies, and reach a high degree of maturity in producing robust solutions in terms of *proof of value* and *proof of concept*.

These concepts have been implemented in a collaborative platform (Rianantsoa *et al.* 2011a, Rianantsoa *et al.* 2011b). Libraries of Airbus in-house *value dimensions*, *value drivers*, *value strategies*, etc, have been defined for each language object of PSK-value model (see Figure 10). Value metrics have been proposed in (Rianantsoa *et al.* 2011a, Rianantsoa *et al.* 2011b) to provide value oriented decision criteria like *contribution to global value*, *contribution to differentiation*, *distance between a strategy and a solution*, etc. Finally, Airbus company has decided to adopt CtV methodology on every new strategic aircraft development project.

To conclude, the CtV methodology allows ensuring an agile management and alignment of the business strategies of an organization and the architectural solutions of its products to be introduced on the market. It supports the collaboration between the business teams, on the one hand, which work on the product planning, and the design engineering teams, on the other hand, which deploy the conceptual design. It permits then to increase and validate both the *proof of value* of solution architectures and the *proof of concept* of business strategies. This collaboration is enabled in a process of *knowledge* capturing, sharing and analysis, as well as in a process of common *value strategies* definition. Our intention of systematic and integrated steering of the conceptual design and product planning stages gives a new insight both for Innovation Marketing and Design Engineering domains.

## 9. Conclusion

We set a goal to develop an innovation engineering to professionalize as much as possible, an innovation supply chain in companies (see section 2), that is to say to ensure a continuous production of innovations. This is obviously inconsistent with the known and conventional principles of production, quality and risk management for one should not "kill innovation" by framing too tight, organizing too much. But one must make possible to deliver a stream of innovations in accordance with the concerned disciplines in companies as well as with the layers of business processes that contribute to it (see Figure 3).

We are also at the meeting of several scientific disciplines that have their language, their methods of observation, production and model validation. Let us mention: design engineering, industrial engineering, strategic management of innovative projects, management of technologies, strategic marketing, business information systems, process modeling, accounting and market economy.

While attempting an open vision of contributing disciplines, we must propose design models and methodologies to be based on theories, implemented in the most relevant manner (algorithm, database, platform, business processes ), deployed at an adequate scale in companies in the most ingenious way and validated as rigorously as possible: qualitatively and/or quantitatively (see (see (Yannou and Petiot 2011) and (Blessing and Chakrabarti 2009)).

We have shown in this paper that it could be possible to develop new models and new management methodologies for innovation processes and projects in the context of companies; the methodological framework *Radical Innovation Design*® is an example.

## 10. References

Astebro T., 2004. Key Success Factors for Technological Enterpreneurs' R&D Projects. *IEEE Transactions on Engineering Management*, 51 (3), 314-321.

Aurisicchio M., Bracewell R., 2009. *Engineering design by integrated diagrams*, *In ICED*, Stanford University.

Aurisicchio M., Gourtovaia M., Bracewell R., Wallace K., 2008. How to evaluate reading and interpretation of differently structured engineering design rationales. *Artificial Intelligence for Engineering Design, Analysis and Manufacturing*, 22, 345-358.

Bertoluci G., Yannou B., Attias D., Vallet E., 2013. A categorization of innovation funnels of companies as a way to better make conscious agility and permeability of innovation processes, *In ICORD: 4th International Conference on Research into Design*, January 7-9, Chennai, India.

Blessing L., Chakrabarti A., 2009. *DRM, a Design Research Methodology* Srpinger, ISBN 978-1-84882-586-4.

Boujut J.F., Blanco E., 2003. *Intermediary Objects as a Means to Foster Cooperation in Engineering Design*. *Journal of CSCW*, 12 (2).

Bracewell R., Wallace K., Moss M., Knott D., 2009. Capturing design rationale. *Computer-Aided Design*, 41, 173-186.

Cuisinier C., Vallet E., Bertoluci G., Attias D., Yannou B., 2012. *Un nouveau regard sur l'innovation - Un état des pratiques et des modèles organisationnels dans les grandes entreprises*, Paris: Techniques de l'Ingénieur, ISBN 978-2-85059-130-3.

Dickson P., 1982. Person-Situation: Segmentation's Missing Link. *Journal of Marketing Research*, 46 (4), 56-64.

Hatchuel A., Weil B., 2003. A new approach of innovative design: an introduction to C-K theory, *In 14th International Conference on Engineering Design - ICED'03*.

He L., Chen W., Hoyle C., Yannou B., 2012. Choice modeling for usage context-based design. *Journal of Mechanical Design, DOI: 10.1115/1.4005860*.

Jeantet A., 1998. Les objets intermédiaires de la conception. *Sociologie du travail*, 3/98, 291-316.

Kim C.W., Mauborgne R., 2005. *Blue ocean strategy - How to create uncontested market space and make the competition irrelevant*, Boston, USA/MA: Harvard Business School press.

Millier P., 1999. *Marketing The Unknown: Developing Market Strategies For Technical Innovations*, New-York: John Wiley&sons.

Motte D., Bjärnemo R., Yannou B., 2011a. On the interaction between the engineering design and development process models - Part I: Elaborate Elaborations on the generally accepted process models, *In ICoRD: 3rd International Conference on Research into Design*, January 10-12, Bangalore, India.

Motte D., Bjärnemo R., Yannou B., 2011b. On the interaction between the engineering design and the development process models - Part II: Shortcomings and limitations, *In ICoRD: 3rd International Conference on Research into Design*, January 10-12, Bangalore, India.

Motte D., Yannou B., Bjärnemo R., 2011c. The specificities of radical innovation, *In ICoRD: 3rd International Conference on Research into Design*, January 10-12, Bangalore, India.

Rianantsoa N., 2012. *Strategical and multidisciplinary steering of aeronautical projects on the basis of shared value model and innovation process - Pilotage stratégique et multidisciplinaire de projets aéronautiques, basé sur un modèle de valeur et de processus d'innovation intégré*. Thèse de Doctorat soutenue le 12 juin 2012. Ecole Centrale Paris, Laboratoire Génie Industriel.

Rianantsoa N., Yannou B., Redon R., 2010a. Concept-to-value: Method and tool for value creation in conceptual design, *In IDETC/DAC: ASME International Design Engineering Technical Conferences & Computers and Information in Engineering Conferences / Design Automation Conference*, August 15-18, Montreal, Canada.

Rianantsoa N., Yannou B., Redon R., 2010b. Dynamics of definition and evaluation of value creation strategies and design concepts, *In IDMME Virtual Concept*, October 20-22, Bordeaux, France.

Rianantsoa N., Yannou B., Redon R., 2011a. Steering the value creation in an airplane design project from the business strategies to the architectural concepts, *In ICED 2011*, Copenhagen, Denmark.

Rianantsoa N., Yannou B., Redon R., Monceaux A., 2011b. Definition of value strategies in complex aeronautical projects: steering both product planning and conceptual design stages throughout a unified value model of knowledge, problems and solutions, *In CSDM*, Paris.

Shah J.J., Kulkarni S.V., Vargas-Hernandez N., 2000. Evaluation of Idea Generation Methods for Conceptual Design: Effectiveness Metrics and Design of Experiments. *Journal of Mechanical Design*, 122 (4), 377-385.

Thompson S.C., Paredis C., 2010. An Introduction to Rational Design Theory, *In IDETC/DTM: ASME International Design Engineering Technical Conferences & Computers and Information in Engineering Conferences / Design Theory and Methodology Conference*, August 15-18, Montreal, Canada.

Wang J., 2012. *A usage coverage based approach for assessing product family design - Une méthode d'évaluation de la conception d'une famille de produits basée sur le modèle de couverture d'usages*. Thèse de Doctorat. Ecole Centrale Paris, Laboratoire Génie Industriel.

Wang J., Yannou B., Alizon F., Yvars P.-A., 2012. A Usage Coverage-Based Approach for Assessing Product Family Design. *Engineering With Computers*, DOI 10.1007/s00366-012-0262-1.

Yannou B., Chen W., Wang J., Hoyle C., Drayer M., Rianantsoa N., Alizon F., Mathieu J.-P., 2009. Usage Coverage Model For Choice Modeling: Principles, *In IDETC/DAC: ASME International Design Engineering Technical Conferences & Computers and Information in Engineering Conferences / Design Automation Conference*, August 30 - September 02, San Diego, CA.

Yannou B., Chen W., Yvars P.-A., Hoyle C., 2012a. Set-based design by simulation of usage scenario coverage. *Journal of Engineering Design*.

Yannou B., Jankovic M., Leroy Y., 2011. Empirical verifications of some Radical Innovation Design principles onto the quality of innovative designs, *In ICED 2011*, Copenhagen, Denmark.

Yannou B., Jankovic M., Leroy Y., Okudan Kremer G., 2012b. Observations from radical innovation projects considering the company context. *Journal of Mechanical Design*.

Yannou B., Petiot J.-F., 2011. A View of Design (and JMD): The French Perspective. *Journal of Mechanical Design*, 133 (5 (june 2011)).

Yannou B., Wang J., Yvars P.-A., 2010a. Computation of the usage contexts coverage of a jigsaw with CSP techniques, *In IDETC/DAC: ASME International Design Engineering Technical Conferences & Computers and Information in Engineering Conferences / Design Automation Conference*, August 15-18, Montreal, Canada.

Yannou B., Yvars P.-A., Wang J., 2010b. Simulation of the usage coverage of a given product, *In International Design Conference - Design 2010*, May 17-20, Dubrovnik, Croatia.

Zimmer B., 2012. *Structuration d'un cluster d'innovation : Application aux projets d'innovation dans une grappe d'entreprises en gérontechnologie*. Thèse de Doctorat. Ecole Centrale Paris, Laboratoire Génie Industriel.

Zimmer B., Yannou B., Stal Le Cardinal J., 2012. Proposal of radical innovation project selection model based on proofs of value, innovation and concept, *In International Design Conference - Design 2012*, May 21-24, Dubrovnik, Croatia.